\documentstyle[multicol,aps,epsf]{revtex}

\newcommand{\be}{\begin{equation}}
\newcommand{\bel}[1]{\begin{equation}\label{#1}}
\newcommand{\ee}{\end{equation}}
\newcommand{\bea}{\begin{eqnarray}}
\newcommand{\ba}{\begin{array}}
\newcommand{\eea}{\end{eqnarray}}
\newcommand{\ea}{\end{array}}

\begin{document}

\twocolumn[\hsize\textwidth\columnwidth\hsize\csname@twocolumnfalse%
\endcsname

\title{Statistical Mechanics of an Urban Cross : A Solvable Model}
\author{M.Ebrahim Fouladvand$^{1,3}$ , Masoud Nematollahi$^{2}$ and Farzad
Zolfaghari$^{2}$ }

\address{$1$ Department of Physics, Sharif University of Technology, P.O.
Box 11365-9161, Tehran, Iran.}

\address{$2$ Tehran Traffic Control Company (TTCC), P.O. Box 15836  ,
Tehran,Iran. }

\address{$3$ Institute for Studies in Theoretical Physics and Mathematics
,P.O. Box 19395-5531, Tehran, Iran.}

\date{\today}

\maketitle

\begin{abstract}

We propose a model for the intersection of two urban streets. The traffic
status of the crossroads is controlled by a set of traffic lights which
periodically switch to red and green with a total period of$T$.
Two different types of crossroads are discussed. The first one describes the
intersection of two one-way streets, while the second type models the
intersection of a two-way street with an one-way street. We assume that the
vehicles approach the crossroads with constant rates in time which are taken
as the model parameters. We optimize the
traffic flow
at the crossroads by minimizing  the total waiting time of the vehicles per
cycle of the traffic light. This leads to the determination of the
optimum green-time allocated to each phase.

\end{abstract}
]

Over the past decade, the majority of the studies in the inter-disciplinary
field  " {\it physics of traffic } " have been allocated to the {\it 
highway
traffic } ( for a review see Ref. \cite {css99,tgf97,tgf99} ) and
less attention has been paid to the urban traffic. The simulation
of traffic flow in a large-sized city is a formidable task and many
degrees of freedom have to be involved (see e.g. \cite{book}). 
In practice, streets of a city form a network of junctions that are
linked togetether.  Each crossroads receives
demands (vehicles
attempting to pass the cross ) and at each crossroads there exists a traffic
light which, with some certain programming, controls the transportation. 
The first model introduced by statistical physicists for the description of
 the city network which is known as the BML model in the literature, uses
a
deterministic cellular automata framework \cite{bml} and predicts a
sort of phase transition from free-flow to a jammed state, but the
model assumptions are too simple to be taken seriously for practical
purposes. Later, a more serious model was introduced by Schadschneider
and
Chowdhury \cite{schad-deb}. This model combines
the BML mode together with the Nagel-Schreckenberg model of highway
traffic \cite{NS}. Despite its nice formulation, there are still a lot of
simplifications which prevent it from being an {\it effective } and
applicable model to a city network.
The empirical mechanism by which the traffic lights are controlled is
generally
divided into two distinct methods : fixed time and real-time.
In the fixed-time method, a fixed value of time is  allocated to the 
traffic light
 as well as its sub-phase times. 
 In the real-time method, which is becoming increasingly popular in great
cities, the
ensemble of crossroads are intelligently controlled by a central
controller.
The control mechanism  is usually based on the concept of
producing a kind of {\it green waves } between the crossroads.
 These waves interact with each other, and, if the passing demands are 
of high values, the green waves may have destructive effects on each
other, and, hence, the concept of green wave may fail to be the final
solution for optimizing the overall flow. \\
In this paper we aim to analyze a single
crossroads in detail in order to find a better insight to the problem 
of optimizing the total flow in cities.
Our single crossroads is the result of the intersection of two perpendicular
 streets. In
their simplest structure, these streets can each direct a one-way traffic
flow. With no loss of generality, we take them as one-way South to North
(S-N)
and West to East (W-E) streets. Cars arrive at the south and the west 
entrances of the crossroads. Here we assume the arrival rates of the cars
, i.e. the number of cars reaching the crossroads per
second, are
constant in time. Although  from everyday driving experiences in cities,
we know that these rates have fluctuations in the course of time; yet
in definite time intervals , the assumption of the constant arrival
rates could be justified at least on a average level. We take the arrival
rates to be $\alpha _1$ (for S-N cars ) and $\alpha _2$ (for the W-E
cars) respectively. Also we denote the passing-rate of cars ( number of
cars passing the crossroads in the unit of time during the green-phase ) by
$\beta_1$ and $\beta_2$. The period of the traffic lights is taken to be
a definite value $T$ which is assumed to remain constant in the time
interval under consideration. The starting time of each cycle of the
traffic light is at the
moment which the light turns green for the S-N street. The S-N light
remains green until $T_1$. At $T_1$ the traffic lights turn
red for the S-N street and simultaneously changes to green for the
W-E street. This is the beginning of the second phase which continues from
$T_1$ to $T$ (end of one cycle). During Phase I ( $ 0 \leq t \leq T_1$
), the S-N cars can pass the crossroads northward and W-E cars are stopped for
the red light.  Over Phase II ( $T_1 \leq t \leq T$ ), the S-N cars
must wait behind
the red light whilst the W-E cars are passing through the crossroads.
 
Now the basic question is how should traffic engineers adjust the
value of $T_1$ in order to optimize the traffic flow throughout the
 intersection?
 Regarding the fact that drivers are identical human
beings and with the assumption that the number of passengers of each
car takes
an equal average value for each direction, the optimization task is
realized by minimizing
the total waiting time of the cars per cycle of the traffic lights. For
this
purpose, we introduce two quantities $N_1$ and $N_2$ which represent the
number of cars stopping at the red lights in the red phases of the S-N
and W-E streets respectively. Clearly $N_1$ and $N_2$ are functions of
time and, in general, are divided into different lanes on the
streets. For the first cycle of the steady time-interval ( in which the
arrival rate are taken to be constant ) the following equations are easily
written down:
\be
N_1(T_1)=[N_1(0) + (\alpha_1 - \beta_1)T_1] \theta
\ee
\be
N_2(T_1)=N_2(0)\theta + \alpha_2T_1
\ee
\be
N_1(T)=[N_1(0) + (\alpha_1 - \beta_1)T_1] \theta +\alpha_1(T-T_1)
\ee
\be
N_2(T)=[N_2(0)\theta + \alpha_2 T -\beta_2(T-T_1)] \theta
\ee

The $\theta$ symbols ensures the positiveness of the quantities
in the brackets, i.e. the value of $\theta$ is one if the quantity in the
bracket is positive and zero elsewhere. This limitation is dictated to us
since, by definition, the quantities $N_1$ and $N_2$ can only take
positive values.

We now define the Total Waiting Time (TWT) of the vehicles per cycle of
the traffic lights. It is the total time wasted by the vehicles during
their
stop in the red phases. The TWT is sum of the sub-waiting
times of each direction. Denoting the TWT and the sub waiting times by
$T^W$, $T^{(W,1)}$ and $T^{(W,2)}$ respectively, the following equations
could
be written for the $n$-th cycle
\be
T_{n \rightarrow n+1}^{(W,1)}= {1\over
2}\alpha _1 (T-T_1)^2 + N_1(nT+T_1)(T-T_1)
\ee
\be
T_{n\rightarrow n+1}^{(W,2)} = N_2(nT)T_1
+ {1
\over 2} \alpha_2T_1^2
\ee
In the
above equations, one must take into account the fact that
during the red phase of a street, not all the cars reaching the crossroads
 wait equally. The cars reaching the intersection just after
the light goes red ( unlucky drivers) have the most waiting time, while
those reaching the crossroads just before the lights turns green (lucky drivers)
spend no time waiting. This leads to an average factor ${1 \over 2}$ in
the above equations.
The value of $N_2(0)$ refers to the number of the stopped cars in the W-E
street just after the instant when the light turns red for the W-E cars 
(in the first cycle).
Here two distinct situations are identified. In the first case, $N_2(0)$
is
nearly zero which describes the situation when the W-E street is
not crowded. The other case is $N_2(0) \geq 0 $ which corresponds to the
situation of heavy traffic in the W-E street. This means that even in the
final stages of the green phase for the W-E direction, still there is a
considerable
number of W-E cars desiring to pass the cross eastwards.The same scenario
goes with the value of $N_1(T_1)$, and, as we will see in
what follows, these different cases lead to different prescriptions of the
optimization. Before investigating these cases in details, it should be mentioned 
that for the arbitrary values of the rates we do not have a
periodical structure in the values of the $N_1$ and $N_2$. This can be
easily verified by comparing $N_1(T)$ and $N_2(T)$ with their values
at
$t=0$ which implies that, in general, one has $N_i(0) \neq N_i(T)$ for
$i=1,2 $. Now let us discuss the situation where $N_1(T_1)=N_2(0)=0. $
This refers to a light traffic flow in both directions during
the first cycle of the interval. One can easily check that if the 
following stability conditions 
 $\alpha_2 T -\beta_2(T-T_1) \leq 0 $ and $ \alpha_1 T - \beta_1T_1 \leq 0
$ hold, Then we always have the similar conditions in the next cycles
 $N_1(nT+T_1)=N_2(nT)=0$ where $n$ denotes the number of the traffic light
cycle. 
In this particular case which here after will be called State I, the TWT
is
independent of the cycle-number, and minimizing the TWT with respect to
$T_1$ leads to the following equation:
\be
T_1= {\alpha_1 \over \alpha_1 + \alpha _2} T
\ee
this answer must be consistent with the stability conditions which leads
to
the constraints $ \beta_2 \geq \alpha_1 + \alpha_2 $ and $ \beta_1\geq
\alpha_1 + \alpha_2 $ among the rates. We now investigate a totally
different situation i.e. a crowded crossroads. 
Supposing that the first cycle
is characterized by the conditions $ N_2(0) , N_1(T_1) > 0 $. one could
easily verify that provided the relations $ \alpha_1 > \beta_1$ and $
\alpha_2 T > \beta _2 (T-T_1) $ hold, we have a quasi-stationary
condition in the next cycles i.e. $ N_1(nT+T_1), N_2(nT) > 0 $. In sharp
contrast to the State I, (this state which will be referred to as the
State II), the values of $N_1$ and $N_2$ would be functions of the cycle
number. It can be shown that in the large cycle-number limit one has $T_{n
\rightarrow n+1}^{(W,1)} \sim n(T-T_1)(\alpha_1 T-\beta_1 T_1)$ and
$T_{n\rightarrow n+1}^{(W,2)} \sim nT_1(\alpha_2 T -\beta_2 (T-T_1) )$. We
now
minimize the TWT with respect to $T_1$ which leads to the following
equation:
\be
T_1={ \beta_1 +\beta_2 + \alpha_1 -\alpha_2 \over 2(\beta_1 + \beta_2) } T
\ee
the consistency of this solution with stability conditions yields the
following constraints: $ \alpha_1 >\beta_1$ and $\alpha_1 \beta_2 +
2\alpha_2 \beta_1 +\alpha_2 \beta_2 \geq \beta_1 \beta_2 + \beta_2^2$.
Also positiveness of $T_1$ itself imposes the extra restriction $
\alpha_1-\alpha_2 < \beta_1 + \beta_2 $. Next we consider the situation
(State III) where in the first cycle of the traffic light, one has
$N_1(T_1)=0$ but
$N_2(0) > 0$ this correspond to the situation in which the S-N street has
a light traffic flow while the W-E street has a heavy one. The conditions for
having these situations repeated in the next cycles are
$\alpha_1T-\beta_1 T_1\leq 0$ and $ \alpha_2T-\beta_2(T-T_1) \geq 0 $. In
the large cycle-number limit, the minimizing of the TWT leads to the value 
\be
T_1= { \beta_2 - \alpha_2 \over 2\beta_2 }T
\ee
It could be easily checked that the above solution is inconsistent with
the stability conditions, and, hence, is not acceptable as an optimization of
traffic light. This inconsistency will remain in the final state 
( State IV ) which is characterized by $N_1(nT+T_1) > 0$ and
$N_2(nT)=0$. In this
case, the solution obtained by minimizing the TWT is again inconsistent
with the stability conditions. In these last two states, one must
numerically find the consistent minima.     
We next consider another more usual configuration of a crossroads. Here we let
the vehicles move in the both directions on the S-N street but still the
vehicles in the W-E street are restricted to move eastwards. This situation
describes a one-way to two-way urban intersection. Here each cycle
consists of three phases. In the first phase which lasts for $ 0 \leq t
\leq T_1$, the traffic light is green for the S-N cars and red for the
other two directions. During the second
phase which starts at $T_1$ and finishes at $T_2$, the traffic light is 
green for the N-S cars and red for the other two directions. In 
the final phase, which last for $ T_2 \leq t \leq T $, the traffic light
remains green for the W-E cars and red for the N-S as well as S-N
directions.
The entrance rates are taken to be $\alpha_1, \alpha_2 $ and $\alpha_3$
and we denote the passing rate by $\beta_1, \beta_2$ and $\beta_3$ for
each direction respectively.The starting time of the cycles is chosen
to be the moment
at which the traffic light turns green for the S-N direction. Similar
equations for the Vehicle-Numbers $N_1,N_2$ and $N_3$ could be
written down and one in principle can evaluate the TWT in terms of these
quantities. The
exact form of the TWT strongly depends on the positiveness of the
vehicle-numbers just after the traffic lights goes red for the respective
direction. In the case under consideration, eight different possibilities
can occur. Let us only discuss the most probable one which corresponds to
the the following situation: $N_1(nT+T_1) = N_2(nT+T_2) = N_3(nT) =0$ 
In this case the traffic status is light in each of the three directions.
Following the same discussion made for the previous crossroads, it could be
easily verified that the triple stability condition for the validity of
the above assumptions are as follows: $ \alpha_1 T - \beta_1 T_1 \leq 0
$, $ \alpha_2 T -\beta_2 (T_2-T_1) \leq 0$ and $\alpha_3T-\beta_3(T-T_2)$.
It can be shown that minimizing the TWT with respect to $T_1$ and $T_2$

leads to the following fixation of $T_1 , T_2$
\be
T_1 = T{ \alpha_1(\alpha_2 + \alpha_3) -\alpha_2\alpha_3 \over
 \alpha_1(\alpha_2 + \alpha_3) + \alpha_2\alpha_3 }
\ee
\be
T_2 = T{ 2\alpha_1 \alpha_2 \over
 \alpha_1(\alpha_2 + \alpha_3) + \alpha_2\alpha_3 }
\ee

One directly observes that in the symmetric case of equal arrival rates,
 $T_1$ and $T_2$ take the expected values ${T \over 3}$ and ${ 2T
\over 3}$ respectively. The other point which must be mentioned is that in this
 traffic status, $T_1$ and $T_2$ do not depend on the passing rates,
and are solely determined by the arrival rates. The remaining seven
status will be discussed elsewhere.\\
For comparison of our model to the empirical data, a time-series analysis
on two of Tehran intersections were carried out. The central part of
Tehran is under control of the SCATS ({\it Sydney Coordinated Adaptive
Traffic System}) that is an intelligent
traffic controller system. The strategy followed by most of urban traffic
control systems is based on establishing green-wave along the major streets
of cities. We considered two different intersections. The first one which
is
located in Tehran downtown connects Valiasr St. to Takhtejamshid St. Both
of these  are one-way and major streets. The other crossroads
connects Abbasabad St.(one-way) to Mahnaz St.(one-way) .In contrast to
the previous case, here the first street is a major while the second
street is a
minor one. The data set is provided by magnetic counting loops which
are installed just before the pedestrian-line of each crossroads. The data
were collected on second of July, 2000. The three following figures show a
time-series of data for our major-to-minor crossroads.

\begin{figure}\label{Fig1}
\epsfxsize=4.3truecm
\centerline{\epsfbox{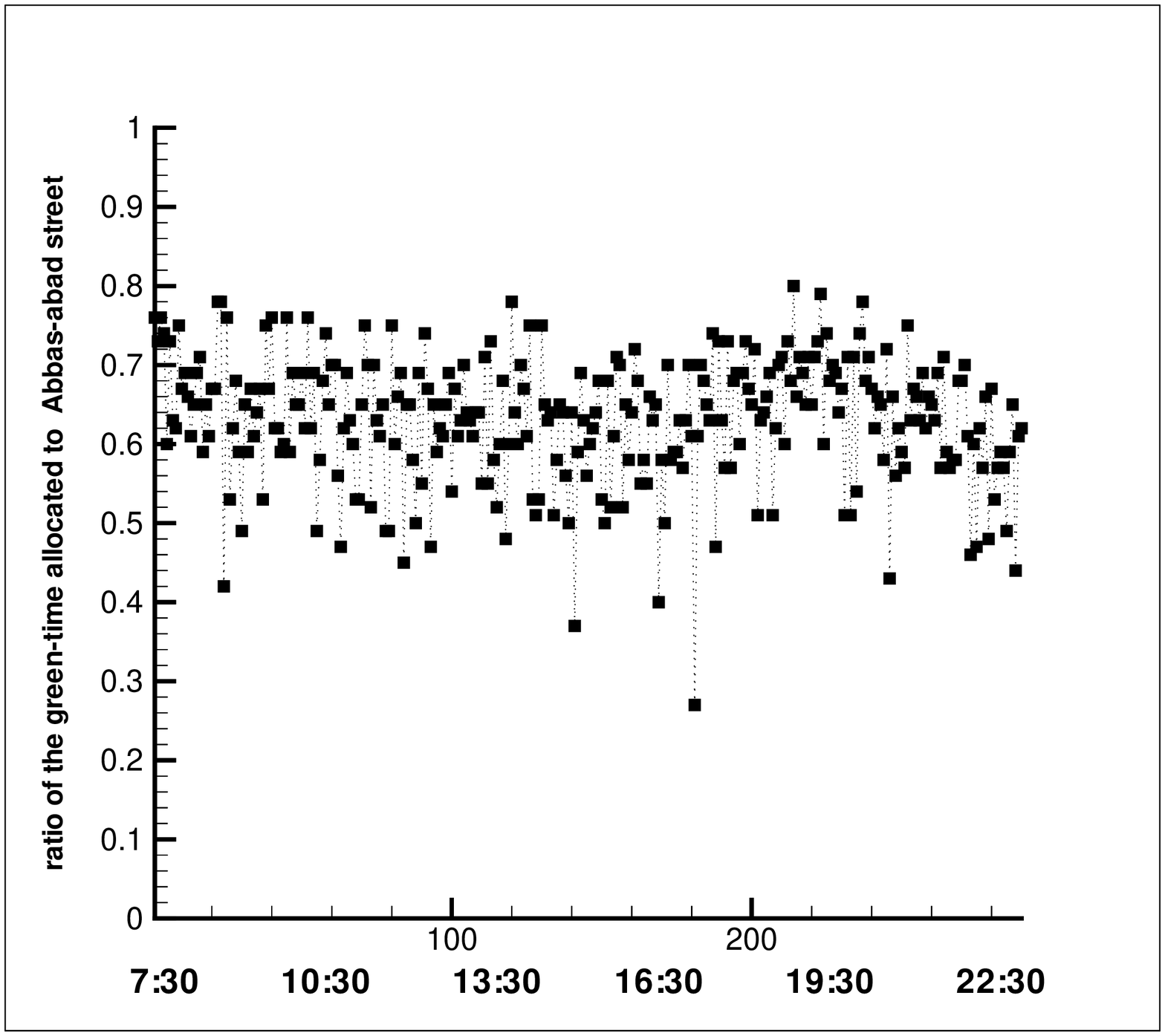}}
\end{figure}
\vspace{0.05 cm}
{\small{Fig.~1: The cycle-ratio of green-time of Abbasabad street to the
total
period of the traffic light cycle. The numbers on the horizontal axes are
the cycle number. } }
 
\begin{figure}\label{Fig2}
\epsfxsize=4.3truecm
\centerline{\epsfbox{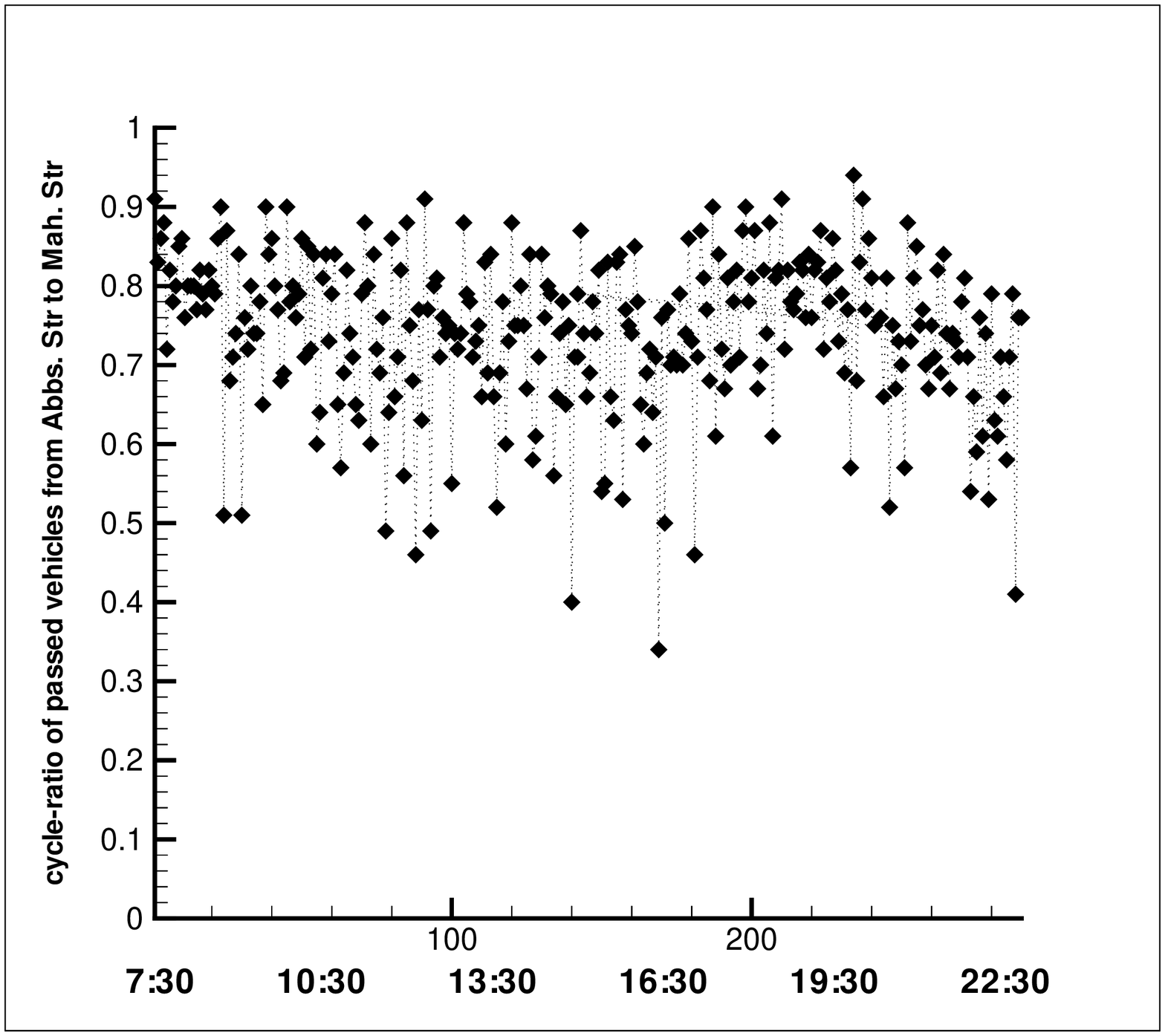}}
\end{figure}
\vspace{0.05 cm}
{\small{Fig.~2: The cycle-ratio of the passed vehicles (during the green
phases) of Abassabad street to Mahnaz street of each traffic light cycle. 
The numbers on the horizontal axis denote the cycle number.} }
 
\begin{figure}\label{Fig3}
\epsfxsize=4.3truecm
\centerline{\epsfbox{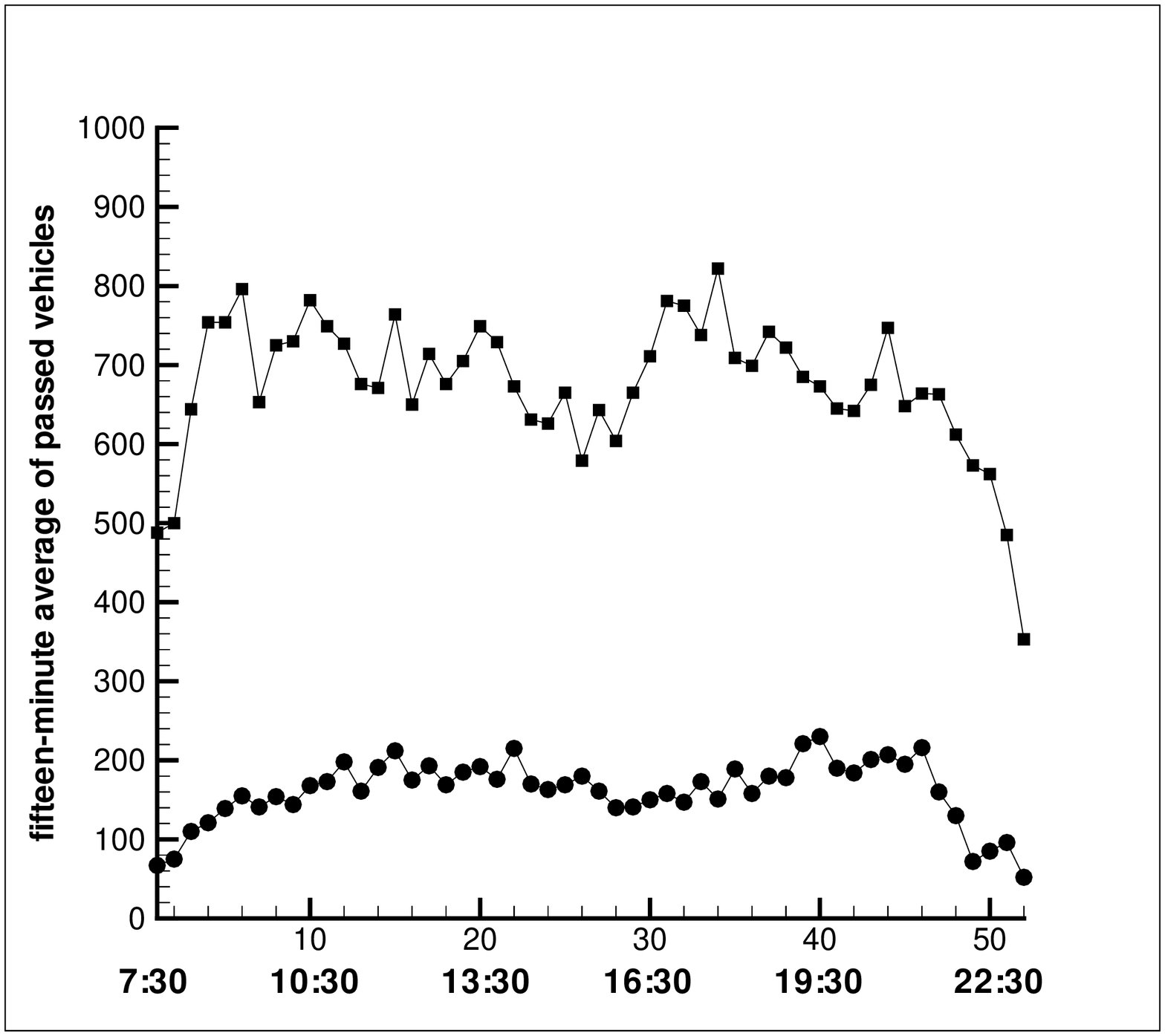}}
\end{figure}
\vspace{0.05 cm}
{\small{Fig.~3: the above graph denotes the total number of
passed vehicles (from both streets) during fifteen-minute time
intervals. The lower graph shows the number of passed vehicles form
Mahnaz street during fifteen-minute intervals. } }\\ 
As seen from the graphs, the cycle-ratio of the green time allocated to
Abbasabad str. strongly fluctuates due to demand fluctuations received
by the cross. In order to have a rough estimation of our model
parameters, we considered a two-hour time interval between
12:30-14:30 . We approximated the ratio of ${ \alpha_1 \over \alpha_2}$ by
the ration of the total number of vehicles which have passed through
Abbasabad  Str. to that number for Mahnaz Street. This yields the value
${\alpha_1 \over \alpha_2 }= 0.35 $. Putting this value into
equation (7) yields ${T_1 \over T}=0.74$. On the other hand the averaged
value of the
cycle-ratio
of the green time of Abbasabad street over the two-hour period leads to
result ${T_1 \over T }=0.64$. The following figures belong to our
major-to-major crossroads. 

\begin{figure}\label{Fig4}
\epsfxsize=4.5truecm
\centerline{\epsfbox{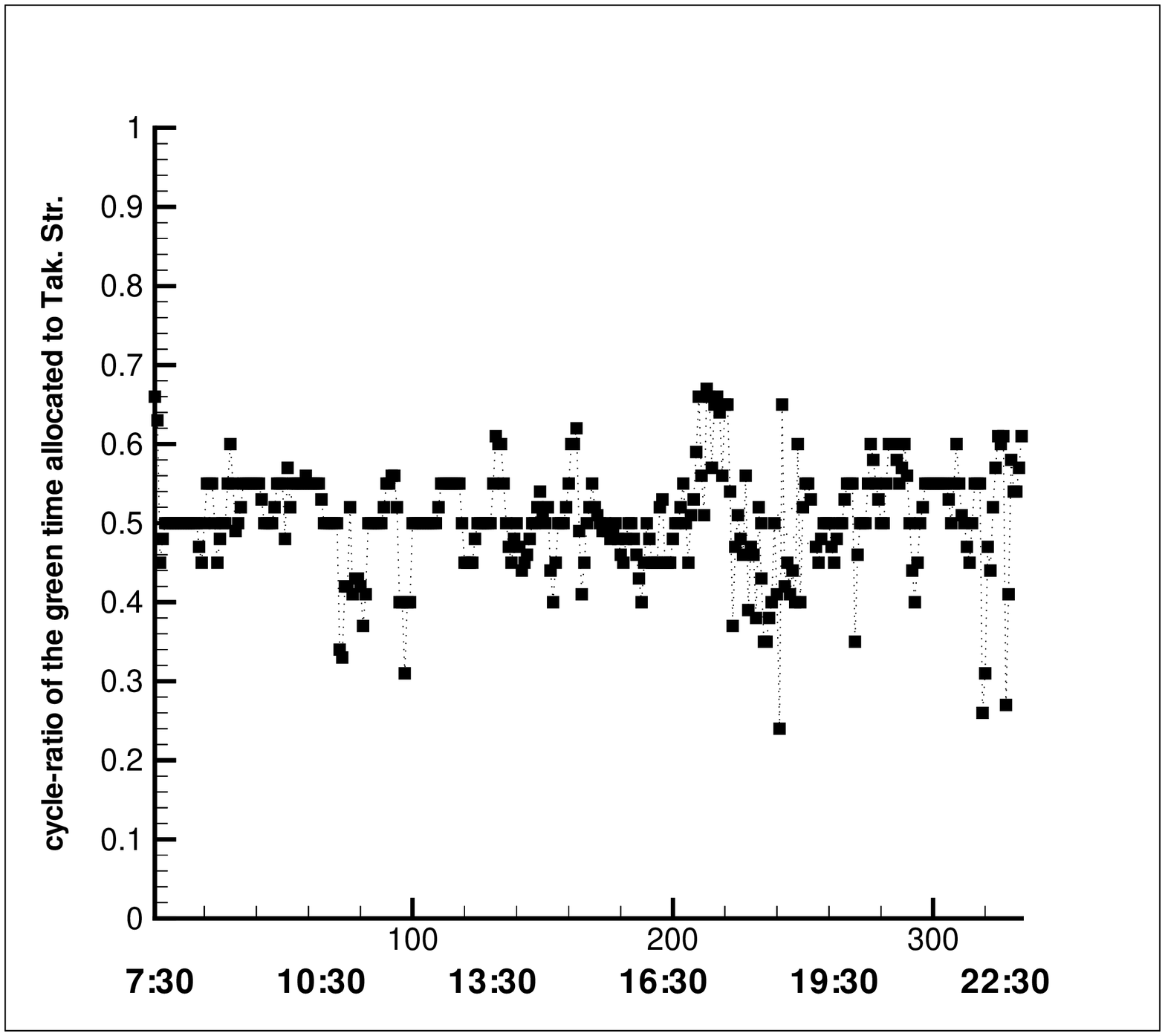}}
\end{figure}
\vspace{0.05 cm}
{\small{Fig.~4: the cycle-ratio of green time allocated to Takhtejamshid
Street. The numbers on the horizontal axis denote the cycle number.  } }\\ 

\begin{figure}\label{Fig5}
\epsfxsize=4.5truecm
\centerline{\epsfbox{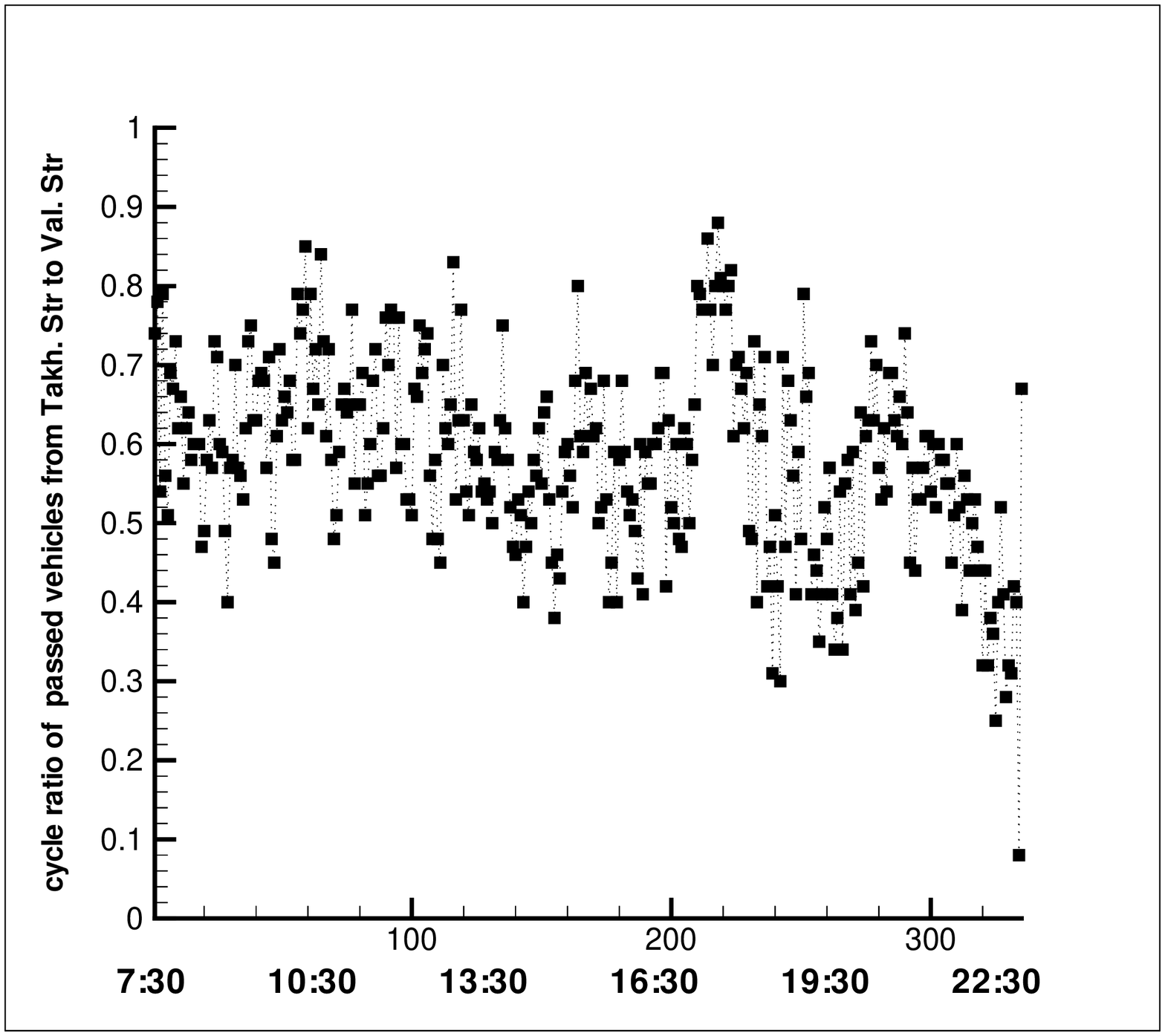}}
\end{figure}
\vspace{0.05 cm}
{\small{Fig.~5:  cycle-ratio of the passed vehicles from Takhtejamshid
street to Valiasr street. } }\\

\begin{figure}\label{Fig6}
\epsfxsize=4.4truecm
\centerline{\epsfbox{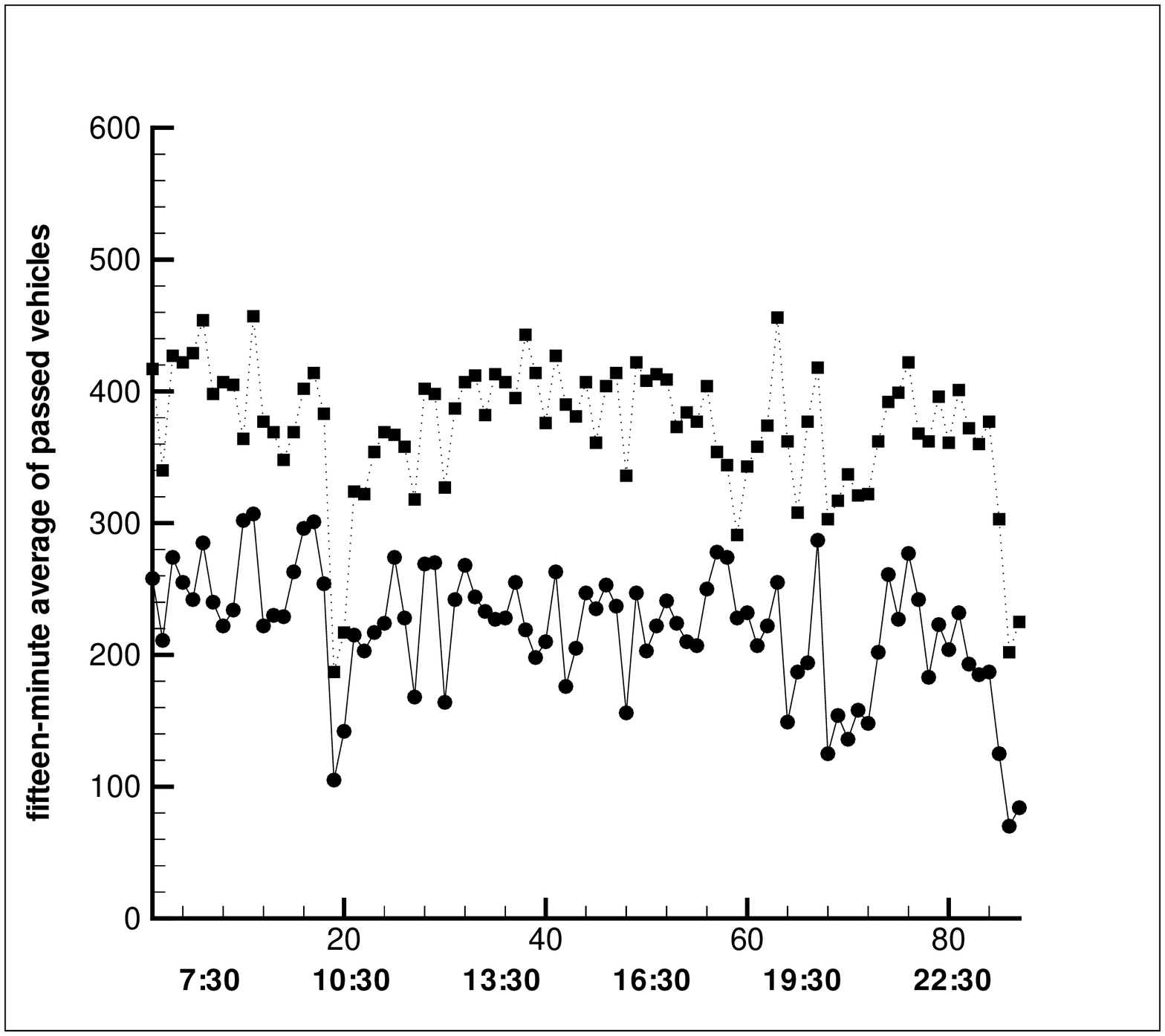}}
\end{figure}
\vspace{0.05 cm}
{\small{Fig.~6: the above graph denotes the total number of
passed vehicles (from both streets) during fifteen-minute time
intervals. The lower graph shows the number of passed vehicles form
Takhtejamshid str. during fifteen-minute intervals. } }\\ 
Here we focused on the interval 13:30 - 15:30. The averaged value of
${T_1 \over T}$ is 0.52 while the same amount evaluated from eq.(7)
is 0.56.

In conclusion, we have developed a prescription for the traffic-light
programming at a single urban crossroads. The method is based on minimizing the
TWT of cars stopping in the red phases of the traffic light. 
The value of $T$ should be
so tuned that during the green phases, time-headway is less than a certain
value.
In fact, in the model the values of the passing rate refer to the maximum
capacity of cross ( maximum number of cars passing the cross in the unit
of time ), which is plausible if the $T$ is appropriately adjusted with 
the congestion of the crossroads. Optimizing the traffic at each crossroads is the
stating point of the more comprehensive problem of city network.
Nevertheless, our model best suits those marginal intersections 
of cities where the effect of the other crossroads is suppressed.
 
{\bf Acknowledgments}: M.E.F is grateful to {\it Tehran Traffic Control
Company } for the data support. We would like to express our gratitude to
R. Sorfleet, M.Arab salmani, M. Salmani and M. Khoshechin for useful
helps. 
\bibliographystyle{unsrt}

\end{document}